\documentclass{aa}  
\pdfoutput=1

\usepackage{graphicx}
\usepackage{txfonts}
\usepackage{indentfirst}
\usepackage{natbib}
\usepackage{ulem}
\bibpunct{(}{)}{;}{a}{}{,} 

\begin{document}
  
  \title{Protostellar collapse: radiative and magnetic feedbacks on small scale fragmentation.}
  
   \author{B. Commer\c con
           \inst{1,2,3}
           ,
           P. Hennebelle\inst{4}
 	  ,
	  E. Audit\inst{3}
	  ,
 	  G. Chabrier\inst{2}
	  \and
	  R. Teyssier\inst{3}
          }
   \offprints{B. Commer\c con}

   \institute{Max Planck Institute for Astronomy, K\"onigstuhl 17, 69117 Heidelberg, Germany\\
              \email{benoit@mpia-hd.mpg.de}
   \and
   	\'Ecole Normale Sup\'erieure de Lyon, CRAL, UMR 5574 CNRS, Universit\'e de Lyon,
46 all\'ee d'Italie, 69364 Lyon Cedex 07, France
         \and 
	 Laboratoire AIM, CEA/DSM - CNRS - Universit\'e Paris Diderot,
IRFU/SAp, 91191 Gif sur Yvette, France 
         \and
             Laboratoire de radioastronomie, UMR 8112 CNRS, \'Ecole Normale Sup\'erieure et Observatoire 
de Paris, 24 rue Lhomond, 75231 Paris Cedex 05, France
             }

   \date{Received xx, 2009; accepted xxx, 2009}

  \abstract{It is established that both radiative transfer and magnetic field have a strong impact on the collapse and the fragmentation of prestellar dense cores, but no consistent calculation exists yet at such scales. }   
  {We perform the first Radiation-Magneto-HydroDynamics numerical calculations at a prestellar core scale.}
  { We present original AMR calculations including magnetic field (in the ideal MHD limit) and radiative transfer, within the Flux Limited Diffusion approximation, of the collapse of a 1 M$_\odot$ dense core. We compare the results with calculations performed with a barotropic EOS. }
      {We show that radiative transfer has an important impact on the collapse and the fragmentation, through the cooling or heating of the gas, 
and is complementary of the magnetic field. A larger field yields a stronger magnetic braking, increasing the accretion rate and thus the effect of the radiative feedback. Even for a strongly magnetized core, where the
      dynamics of the collapse is dominated by the magnetic field, radiative transfer is crucial to determine the temperature and optical depth distributions, two potentially accessible observational diagnostics. A barotropic EOS cannot account for realistic fragmentation. The diffusivity of the numerical scheme, however, is found to strongly affect the output of the collapse, leading eventually to spurious
fragmentation.  }  
       {Both radiative transfer and magnetic field must be included in numerical calculations of star formation to obtain realistic collapse configurations and observable signatures. Nevertheless, the numerical resolution and the robustness of the solver are of prime importance to obtain reliable results.  When using an accurate solver, the fragmentation is found to always remain inhibited by the magnetic field, at least in the ideal MHD limit, even  when radiative transfer is included.}   


\titlerunning{}
\authorrunning{B. Commer\c con et al.}
   \maketitle

\section{Introduction}

Understanding star  formation  is  one  of  the  most  challenging  problems  in
contemporary  astrophysics and numerical calculations provide a useful approach to investigate it thoroughly.   Thanks to the steadily increasing computer performances, numerical
calculations can  integrate more  and more physical processes. Among these latter, the coupling between
matter and  radiation is a major issue. Up to the formation of
the first  Larson core  \citep{Larson_1969}, the accreting gas can freely radiate
into space and is nearly isothermal (optically
thin   regime).   Once   the gas   becomes  dense enough  ($\rho  >   10^{-13}$
g cm$^{-3}$), the radiation is trapped and the gas starts to heat up (optically
thick  regime).   The  transition  between  these two  regimes controls the collapse and the
fragmentation of the cloud.  The  cooler the gas,  the more important the fragmentation.  Another  key issue in star formation is the role of the magnetic field, as dense cores
are observed to exhibit coherent magnetic structures \citep[e.g.][]{Heiles_2005}. 

Several authors have investigated the impact of radiation \citep{Boss_et_al_2000,Whitehouse_Bate_2006,Attwood_2009,Krumholz_07,Bate_2009,Offner_2009} and of magnetic field \citep[e.g.][]{Mellon_Li_2009,Banerjee_Pudritz_2006, Hennebelle_Teyssier_2008,Machida_et_al_2008,Price_Bate_2007,Price_Bate_2008}  on
the star formation process. So far, however, only \cite{Price_Bate_2009} have performed calculations with both magnetic field and radiative feedback, with an SPH method.  Yet, these calculations explore relatively large scales. In  this Letter,  we  present the first full Radiation-Magneto-HydroDynamics
(RMHD) calculations of the protostellar
collapse at small scales. We investigate in detail the impact of both magnetic field (in the limit of ideal MHD) and radiative  feedback on the fragmentation process and on the
launching of outflows. The results  are compared with
the  ones  obtained with  a  barotropic  equation  of state (EOS),  that
crudely mimics the transition from the isothermal to adiabatic regime, in order to assess the importance of a proper treatment of radiative transfer on the collapse.


\begin{figure*}[t]
  \centering
  \includegraphics[width=14.839cm,height=9.2cm]{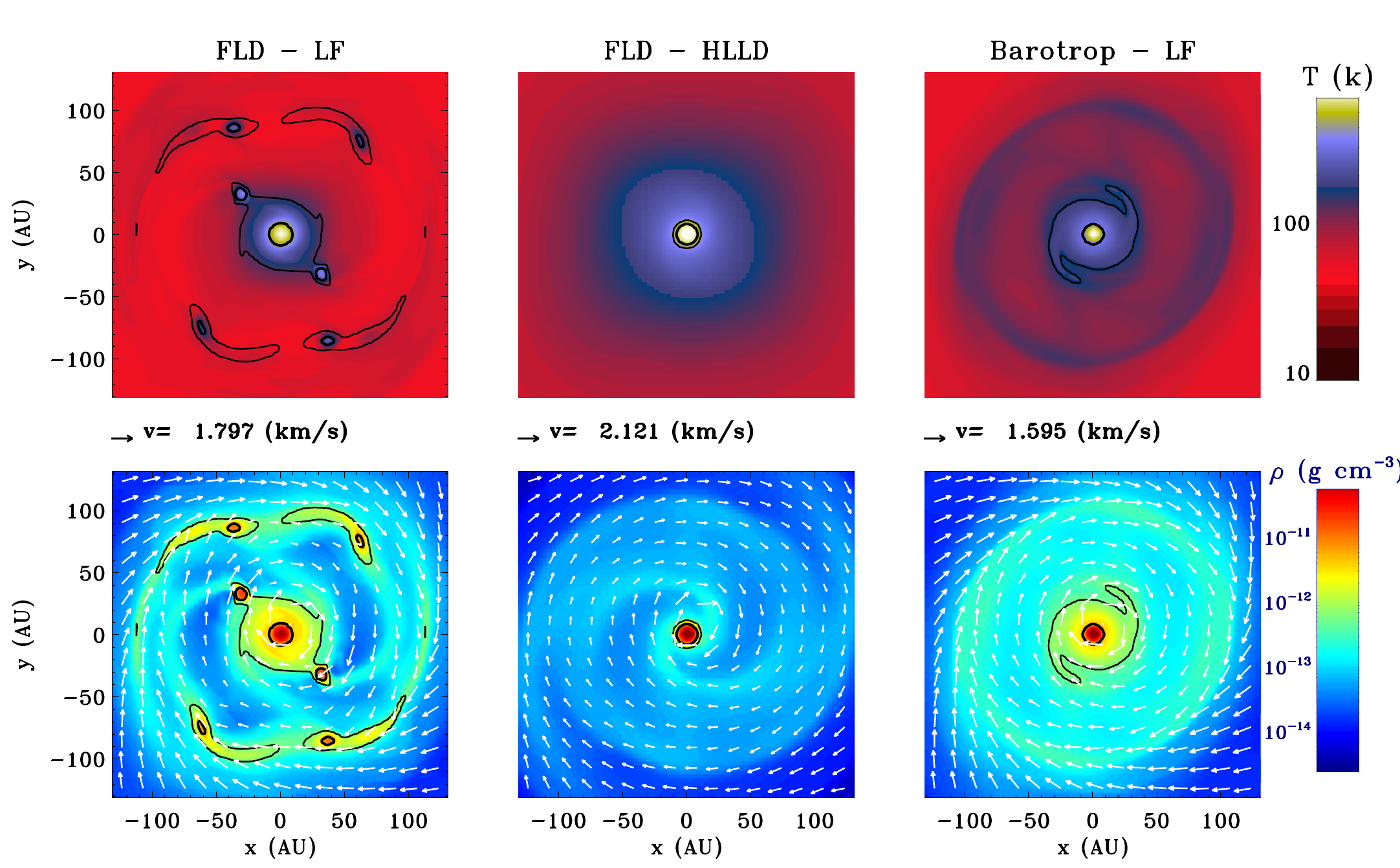}
\caption{Case $\mu=20$: Density and temperature maps in the equatorial plane at time t$=38.6$ kyr: FLD approximation with LF ({\it left}), FLD with HLLD ({\it middle}) and with the barotropic EOS ({\it right}).}
\label{mu20_xy}
\end{figure*}

\section{Numerical method and initial conditions}
We  use the  RAMSES code  \citep{Teyssier-2002} based on a Eulerian formalism (grid based method with Adaptive Mesh Refinement).  
We solve the transfer equations in
the  Flux  Limited Diffusion \citep[FLD, ][]{Minerbo_1978JQSRT} in the  comoving
frame  (fluid
frame) to  evaluate  the  radiative  quantities. The ideal MHD equations  are integrated using an unsplit second-order  Godunov  scheme  
\citep{Fromang_2006}.   Coupling
terms between matter  and radiation, as well as  radiation transport,
are integrated implicitly  in order to handle the  very short heating,
cooling and  diffusion time scales. Details on the numerical method and its implementation will be presented in a forthcoming paper.

Calculations have  been performed using
either the  rather diffusive Lax Friedrich (LF) Riemann solver or the more accurate HLLD Riemann solver \citep{Miyoshi_Kusano_05}. Following up on former studies \citep{Commercon_2008}, we impose at least 10 cells per Jeans length as a grid refinement criterion (parameter $N_\mathrm{J}$). 
 The initial resolution of the grid contains $64^3$ grids.
We  use the low temperature
opacities of \cite{Semenov_et_al_2003A&A}, parametrized as functions of the gas temperature and density.  In contrast, the
non radiative calculations have been performed with a barotropic EOS
$P/\rho = c_{\mathrm{s}0}^2[1+\left(\rho/\rho_\mathrm{ad}\right)^{\gamma -1}]$,
where  $\rho_\mathrm{ad}=2.3 \times 10^{-13}$ g cm$^{-3}$ is  the critical  density at  which  the gas
becomes adiabatic, $\gamma$ (set up to 5/3) is the adiabatic index and $c_{\mathrm{s}0}$ the isothermal sound speed.

We adopt  initial
conditions,   similar   to    those   chosen   in   previous   studies
\citep[e.g.][]{Commercon_2008}.      We     consider    a
uniform-density  sphere  of molecular  gas,  
rotating  about   the  $z$-axis   with  a  uniform   angular  velocity. 
In the present study, the prestellar core mass is fixed at $M_0 =  1$ M$_{\sun}$  and the
temperature at 11 K, which corresponds yo
$c_{\mathrm{s}0}   \sim  0.19   $  km s$^{-1}$.    To promote fragmentation,  we  use  an $\mathrm{m}=2$  azimuthal  density
perturbation with an amplitude of 10\%.
The magnetic field is initially uniform and parallel to the rotation axis. The strength of the magnetic field is expressed in terms of the mass-to-flux over critical mass-to-flux ratio
$\mu=(M_0/\Phi)/(M_0/\Phi)_\mathrm{c}$.
The  initial  energy  balance  is  determined  by  two
dimensionless  parameters, namely  the  ratio of  the
thermal over gravitational energies
$\alpha =0.37$,  and rotational over gravitational energies $\beta=0.045$.
The corresponding free-fall time is $t_\mathrm{ff} \sim 33$ kyr.


\section{Results}

\subsection{Case $\mu=20$\label{mu20}}

We first present the results for $\mu=20$, i.e. a weakly magnetised core. We compare FLD calculations performed with the two Riemann solvers (LF or HLLD) with calculations performed with the barotropic EOS and the  LF solver. The grid refinement criterion is $N_\mathrm{J}=15$.
Performing similar barotropic calculations, \cite{Hennebelle_Teyssier_2008} do not report any fragmentation.

Figure \ref{mu20_xy} portrays temperature ({\it top}) and density ({\it bottom}) maps in the equatorial plane for the three aforementioned calculations at the same time $t\sim 38$ kyr. The black contours represent the transition between the optically thin and thick regions (thin contour) and the quasi adiabatic regions.
The FLD case with the LF solver yields a multiple fragmentation, with a central fragment of mass $5.2\times 10^{-2}$ M$_\odot$ and several (depending on the time and resolution) orbiting fragments of mass $\sim 2-4 \times 10^{-3}$ M$_\odot$ with separations ranging from $\sim 40$ AU to $\sim 100$ AU, whereas no fragmentation occurs in the two other cases. The orbiting fragments in the FLD-LF simulation are quite warm ($\sim 40$ K) while the disk is cold ($\sim 11$ K). In all the simulations, the central fragment ($T > 500$ K) corresponds to an adiabatic region, where the gas cannot radiate away its compressional energy.
In the barotropic case, the mass of the central fragment is $4.9\times 10^{-2}$ M$_\odot$. The outer parts of the disk formed are much denser and warmer than for the two other cases, even though the gas is optically thin and should cool
off efficiently, as illustrated in the FLD-LF calculations. The corresponding values
of the Toomre parameter and Jeans length are larger than in the FLD-LF case. Such heating, however, is spurious and
reflects the approximate treatment of radiative cooling with a barotropic approximation, where temperature is set up by the density.

 The most interesting case is the FLD case with the HLLD Riemann solver. No fragmentation occurs in that case. The matter falls onto a central core of mass $\sim 7.3 \times 10^{-2}$ M$_\odot$. With this less diffusive solver, the generation of a toroidal magnetic field is more efficient. 
The interplay of magnetic field lines and velocity gradients leads  to the effective expansion of a magnetic "tower" in the vertical direction \citep{Hennebelle_Fromang_2008}. The first consequence of this more robust (less diffusive) solver is  that the disc formed is less massive in the equatorial plane than with the LF solver.

\begin{figure}[t]
  \centering
  \includegraphics[width=4.45cm,height=3.3cm]{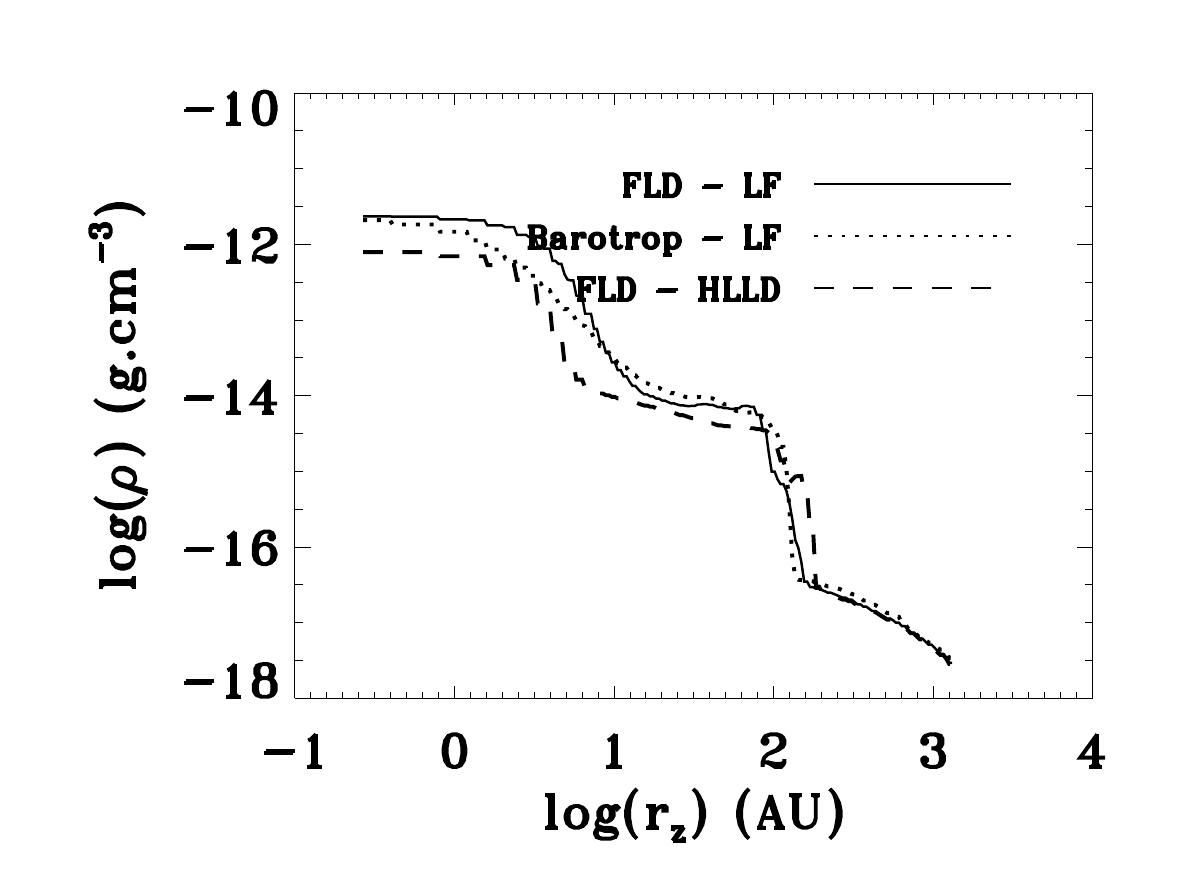}
  \includegraphics[width=4.45cm,height=3.3cm]{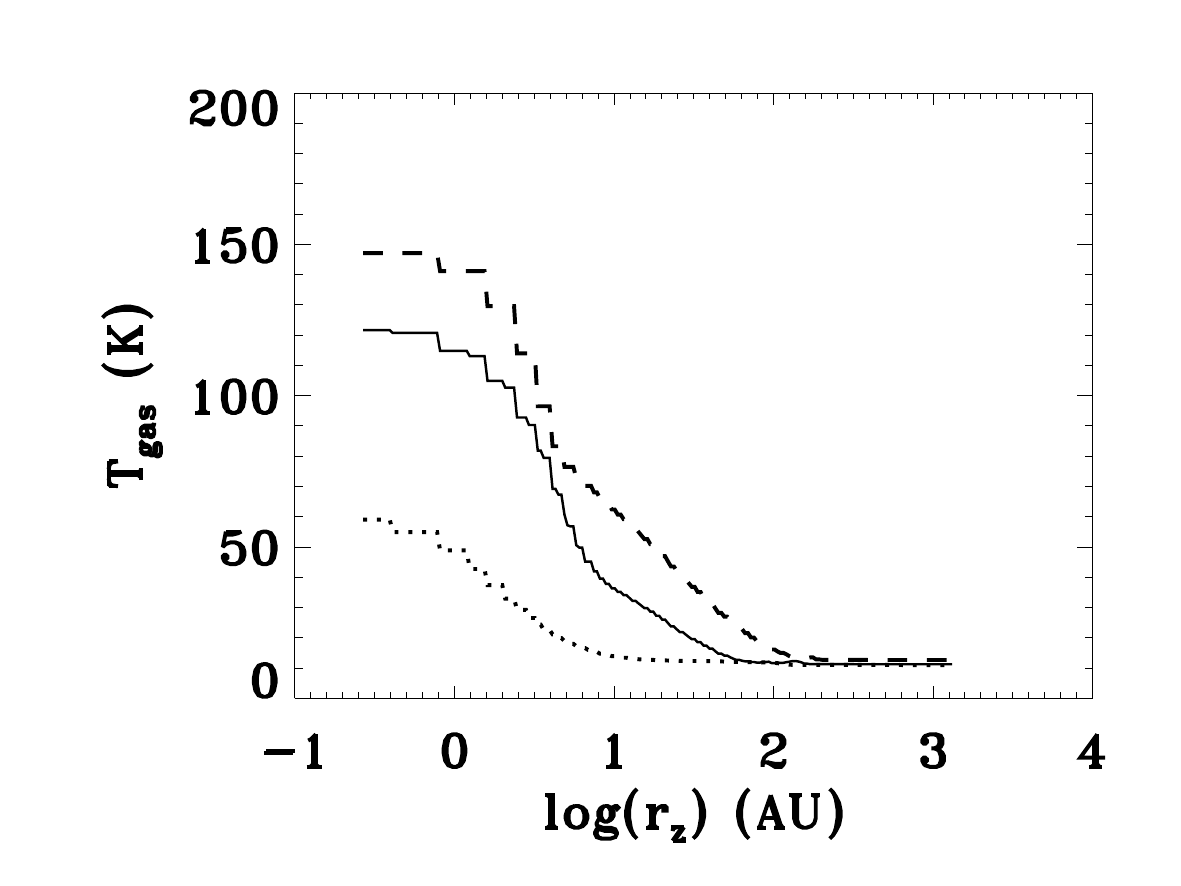}
  \includegraphics[width=4.45cm,height=3.3cm]{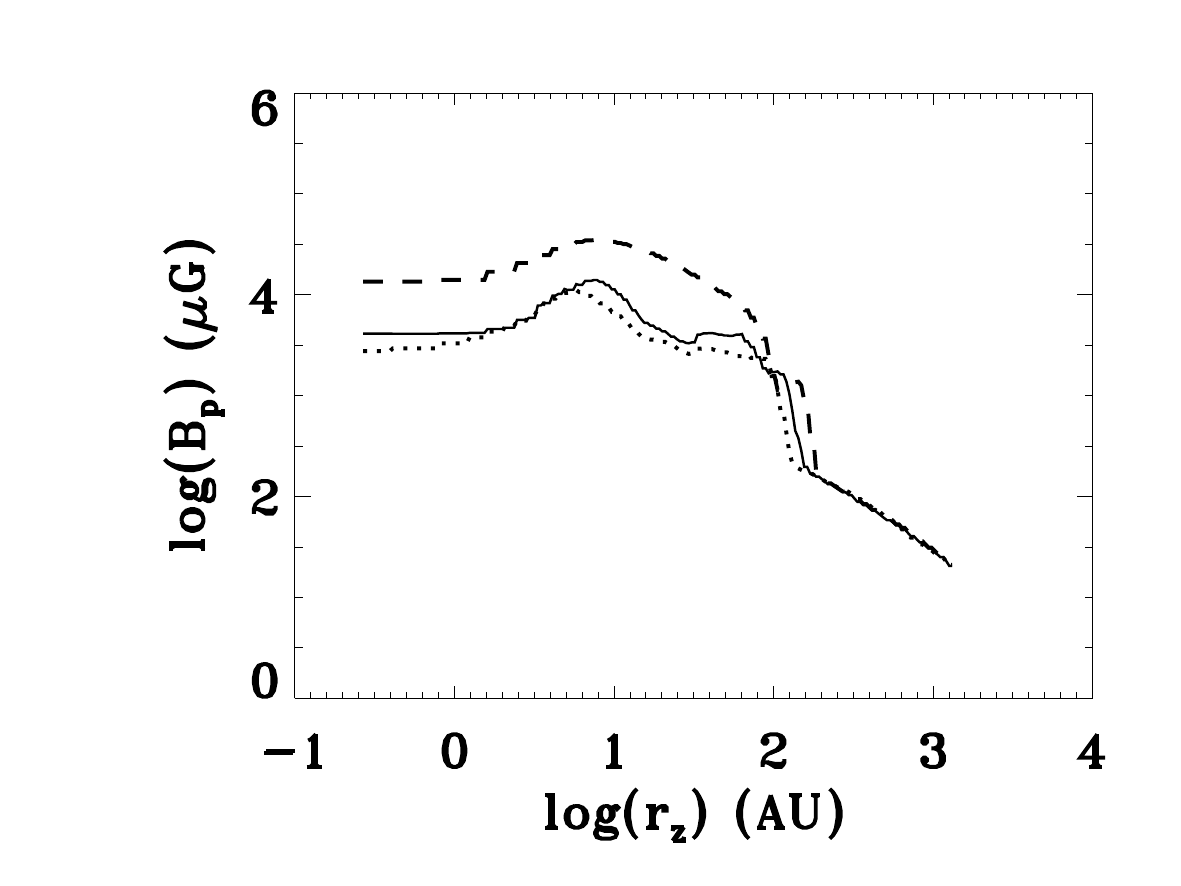}
  \includegraphics[width=4.45cm,height=3.3cm]{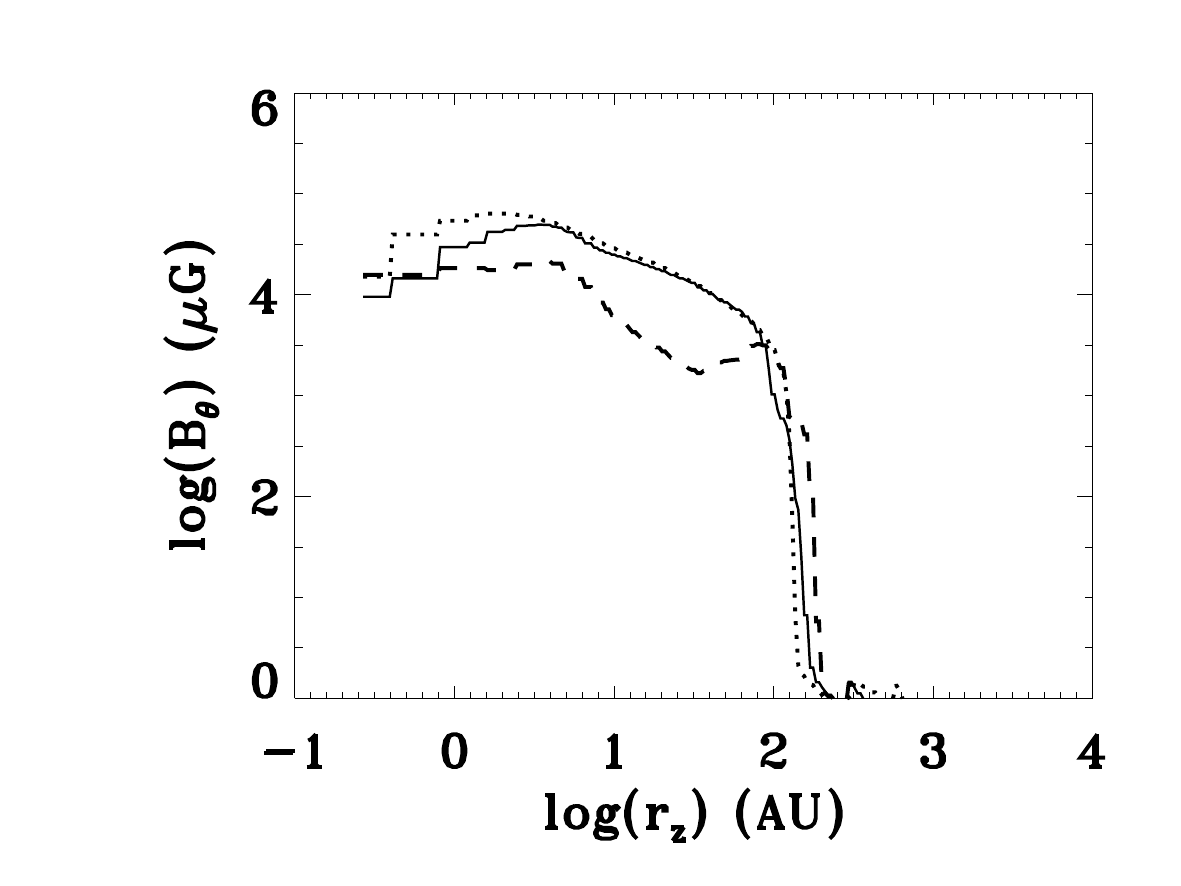}
  \caption{Case $\mu=20$: profiles of density (a), temperature (b), poloidal (c) and toroidal (d) magnetic fields in the $z$-direction at 10 AU from the center, for the same calculations and at the same time as in fig. \ref{mu20_xy}.}
\label{profils_mu20}
\end{figure}

Figure \ref{profils_mu20} shows the density, temperature and poloidal and toroidal magnetic field profiles in the $z$-direction at a distance of 10 AU from the center, for the three aforementioned calculations, at the same time. We have noted that the extent of the magnetic tower depends on (increases with) the resolution. The magnetic tower is less dense close to the equatorial plane, since the gas is spread in the bubble. Contrary to the barotropic case, where the temperature falls to 10 K at a distance $> 10$ AU, the radiative feedback extents vertically up to $\sim 100$ AU with radiative transfer (FLD). 
Although toroidal magnetic field generation is more effective with HLLD than with LF, the toroidal magnetic field in the FLD-LF case is larger at small radii. This is due to the denser disk formed, which compresses and twists the magnetic lines in the region close to the equatorial plane. 
At a distance of 100 AU, the toroidal magnetic field component in the vertical direction is one order of magnitude larger with HLLD than with LF. This favors the extension of the magnetic tower. 
The poloidal magnetic field component is larger everywhere in the HLLD case, since matter is less compressed. 

\begin{figure}[t]
  \centering
  \includegraphics[width=4.45cm,height=3.3cm]{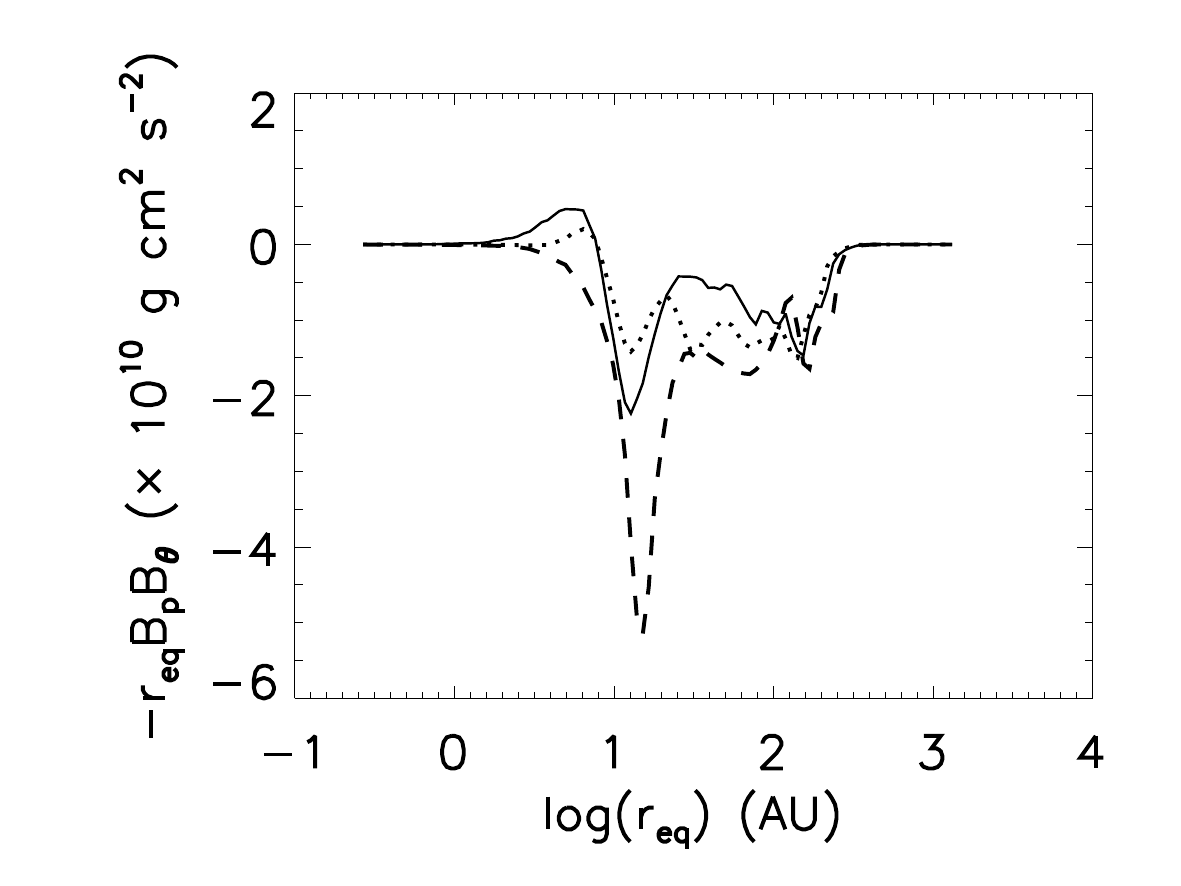}
  \includegraphics[width=4.45cm,height=3.3cm]{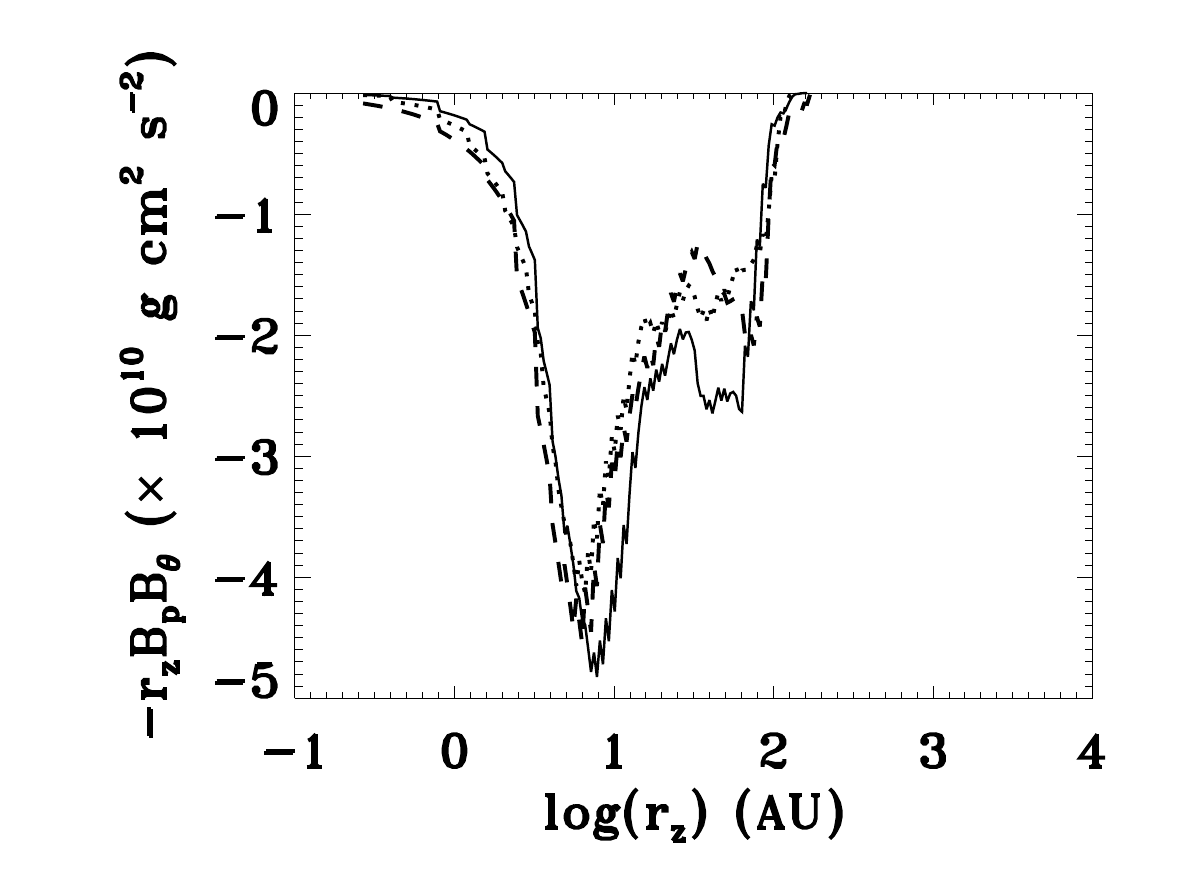}
  \caption{Case $\mu=20$: profiles of magnetic breaking at 10 AU above the equatorial plane (left) and in the $z$-direction (right) at a distance of 10 AU from the center, for the same calculations (same legend) and at the same time as in fig. \ref{mu20_xy}.}
\label{profils_brak}
\end{figure}

Figure \ref{profils_brak} displays the profiles of the magnetic braking at 10 AU above the equatorial plane and in the $z$-direction at a distance of 10 AU from the center, for the three same calculations.
 In the equatorial plane, the magnetic braking obtained with the LF solver or in the barotropic case is barely significant.
Magnetic braking favors faster accretion on the central object and thus yields a larger amount of kinetic energy to be radiated away at the first core accretion shock ({\it all} the in-falling gas kinetic energy is radiated away at the first Larson core accretion shock). The gas is then significantly heated up around the central core. In opposite, there is more angular momentum in the (diffusive) FLD-LF case, which promotes fragmentation at the centrifugal barrier. The accretion in the central fragment is thus smaller and the heating due to accretion is less important. Magnetic braking in the $z$-direction is similar for the three calculations.
We thus identify two important processes, both quantitatively affected by the numerical treatment: the presence of a magnetic tower, whose extension depends on the numerical resolution, and the strong magnetic braking in the equatorial plane, whose strength depends on the diffusivity of the solver.

\begin{figure}[t]
  \centering
  \includegraphics[width=6.6cm,height=4.4cm]{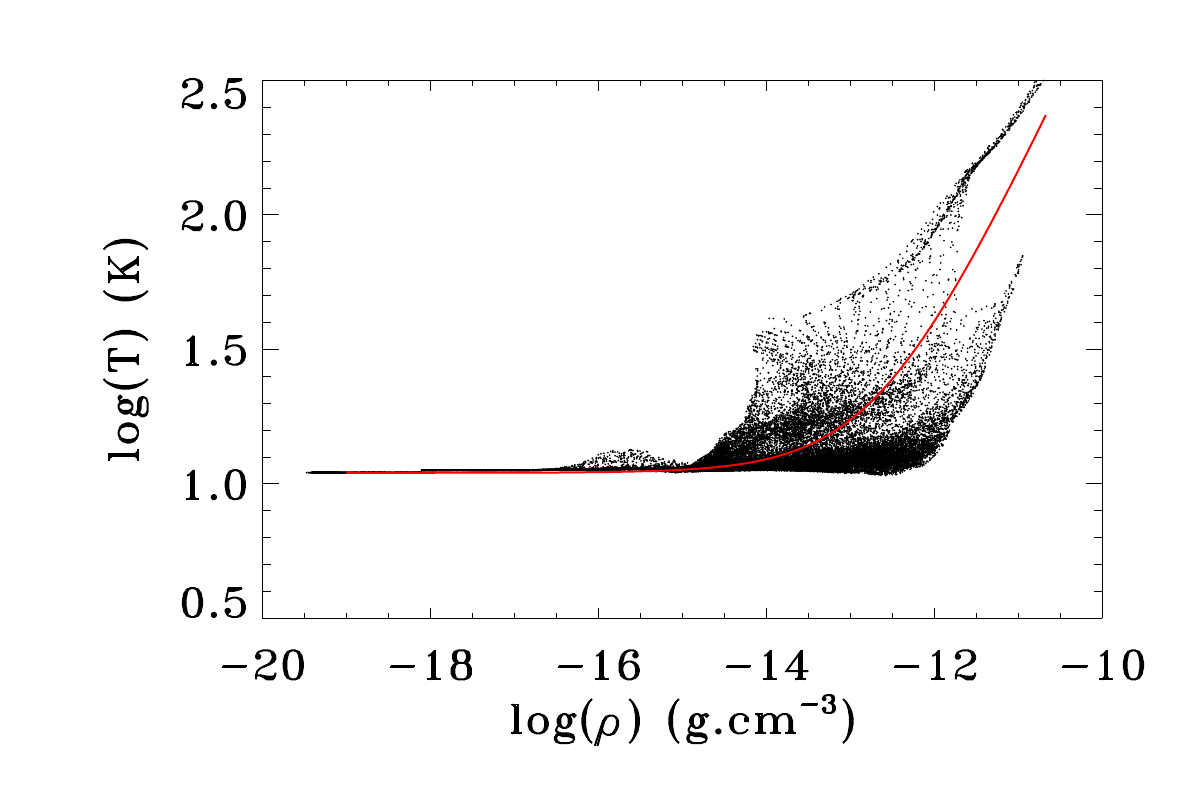}
  \caption{Case $\mu=20$: temperature-density plot for each cell in the FLD-LF calculations (black) and with the
  barotropic  EOS (red line).}
\label{T_rho_mu20}
\end{figure}

Figure \ref{T_rho_mu20} shows the temperature - density distribution for the FLD calculations with the LF solver. The isothermal and  adiabatic regimes  are
  recovered at low and high  density ($\rho < 10^{-16}$ g cm$^{-3}$ and
  $\rho  >  10^{-12.5}$  g cm$^{-3}$).    In  between,  we  observe  a
  dispersion in the (T, $\rho$) plane, where matter can be hotter at low density and cooler at high density compared with the barotropic treatment (red solid line). As seen in the figure, the fragments eventually reach different isentropes. The central fragment lies on the highest entropy level, while the orbiting fragments end up on cooler isentropes. The barotropic EOS cannot reproduce such a spread in temperature at a given density. The entropy and the minimum Jeans mass in that case are set up by the choice of $\rho_\mathrm{ad}$. In the FLD-LF case, cooling is more efficient; material in the equatorial plane cools down by radiating in the vertical directions, where the gas is optically thin. However, this result is mainly affected by the diffusivity of the Riemann solver. 
    In the FLD-HLLD case, all the material has been heated up by the central fragment accretion luminosity  and is distributed above the barotropic adiabat. 
In the FLD-LF case, the fragmentation is the combined result of the
inefficient magnetic braking
and of the efficient cooling due to the FLD. This illustrate the great
importance and the complexity
of the interplay between magnetic field and radiative  transfer.
    

\subsection{Case $\mu=5$}
We now consider a strongly magnetized core, with $\mu=5$, and explore the impact of radiative transfer on the outflow and on the temperature distribution in that case. 
Calculations have been performed with the HLLD solver and $N_\mathrm{J}=10$ . For this more magnetized model, the magnetic field lines determine the dynamics of the collapse and, contrary to the previous case, it is the flow which ajusts to the magnetic field topology rather than the opposite. 
\begin{figure}[t]
  \centering
  \includegraphics[width=8.924cm,height=8.cm]{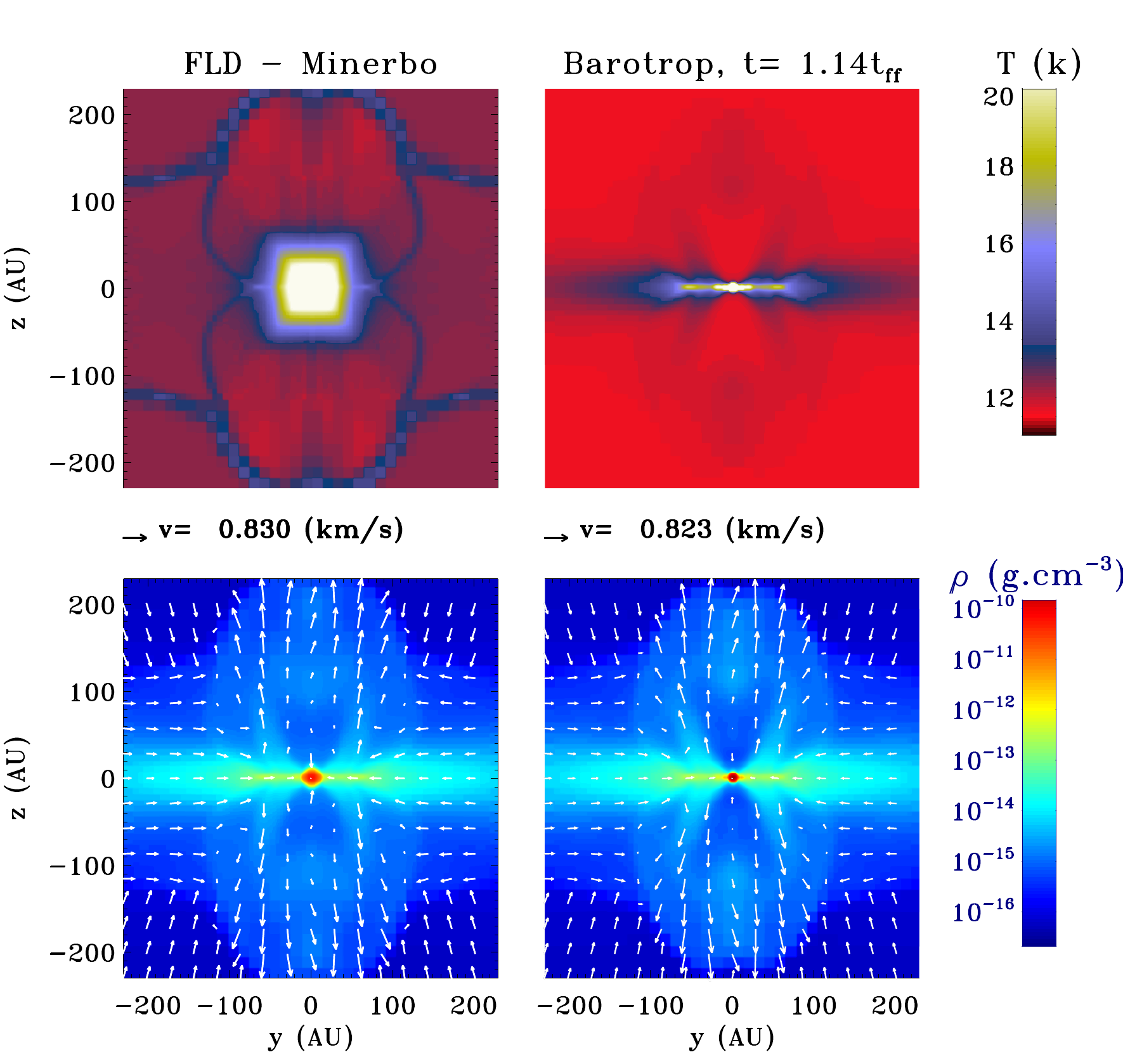}
  \caption{Case $\mu=5$: temperature ({\it top}) and density ({\it bottom}) maps in the $xy$-plane for FLD and barotropic EOS calculations  ({\it right}). }
\label{mu5_yz}
\end{figure}

Figure \ref{mu5_yz} compares temperature and density maps in the $yz$-plane for the FLD and barotropic calculations, at $t=1.14\,t_\mathrm{ff} \sim 38$ kyr. In both cases, an outflow is launched, with similar propagation and opening angles. The density patterns are also similar. The temperature distributions, however, differ drastically. As shown in \cite{Hennebelle_Fromang_2008}, a pseudo warm disk is formed in the equatorial plane in the barotropic case. Note also the lack of significant heating in the vertical direction with the barotropic EOS. Since the density within the outflow is low, $\rho  \sim 5\times 10^{-15}$ g cm$^{-3}$, this implies a low temperature, which quickly falls to 11 K at 10 AU from the center. 
When including a more proper treatment of radiation, with the FLD, the radiation escapes preferentially in the vertical direction (along the $z$-axis) and heats up the gas up to about 100 AU. The highest outflow velocity, $v_\mathrm{r}\sim 1.4$ km s$^{-1}$, is obtained in the barotropic case while $v_\mathrm{r}\sim 1.1$ km s$^{-1}$ in the FLD calculation. The border line between the outflow and the in-falling gas shows a small spike, where the material is shocked and heated. 
The toroidal magnetic field component profiles are in good agreement with the results of \cite{Hennebelle_Fromang_2008} and exhibit a nearly constant plateau in the region of the outflow.
For our present study of a 1 M$_\odot$ core, radiative cooling in the $\mu=5$ case is found to be almost inconsequential on the collapse. The cores are never found to fragment. For more massive cores with stronger radiative feedback, however, the heating in the outflow is likely to be more dynamically significant \citep{Krumholz_07}.


\section{Summary and Discussion}
 In the present Letter, we have explored the effects of both radiative transfer and magnetic fields (in the limit of ideal MHD) on the fragmentation of a 1 M$_\odot$ prestellar core. In agreement with the past non-magnetic studies of, for instance, \cite{Boss_et_al_2000,Whitehouse_Bate_2006, Krumholz_07}, we show that a proper treatment of radiation is important to correctly describe this process. Radiative transfer enables the gas to significantly cool or heat up in different regions of equal densities whereas a barotropic EOS approximation implies that the cooling and the heating are fixed by the density. 

In the case $\mu=20$, where both the behaviour of the flow and of the magnetic field affect the dynamic of the collapse, we show that radiative transfer has an important impact on the final structure. A barotropic approximation cannot account either for the cooling of the dense and rotating gas in the equatorial plane nor for the heating of the less dense gas in the vertical direction, where radiation is found to escape preferentially. Although a value $\mu=20$ is not favored by observations, this case clearly illustrates the impact of a proper treatment of radiative transfer on the collapse and fragmentation of prestellar cores. 
For the strongly magnetized $\mu=5$ case, the dynamics of the collapse is dominated by the magnetic field. In that case, a proper treatment of
radiative cooling is less consequential for the collapse itself but is crucial to derive correct optical depth and temperature distributions, two accessible observational diagnostics. Indeed, the temperature distribution with the FLD is found to strongly depend on the geometry
whereas the barotropic approximation yields a nearly uniform distribution. Note also that in the $\mu=5$ case, we always obtain an outflow, even when using the more diffusive LF Riemann solver. Such outflows can not be obtained presently with SPH calculations. 
We also performed FLD test calculations including the radiative feedback from a pseudo central newborn star and found that, for
our 1 M$_\odot$ core case, it is inconsequential at the length and time scales of interest. 

The spurious diffusivity due to either the numerical resolution or the hydrodynamic solver, however, is found to significantly affect the collapse and
the fragmentation. A less diffusive numerical scheme (HLLD instead of LF) produces strong magnetic braking, that transports angular momentum. Material then falls on the central core, leading to a larger accretion luminosity.
A good numerical resolution is also crucial in the vertical direction to accurately describe the growth of the magnetic tower, which spreads the gas around the central object.

\begin{acknowledgements}
We thank the anonymous referee for comments which have improved the paper significantly. 
Calculations have been performed at CEA on the DAPHPC cluster. 
We acknowledge funding from the European Community via the P7/2007-2013 Grant Agreement no. 247060. 
\end{acknowledgements}
\bibliographystyle{aa}
\bibliography{Commercon_letter}


\begin{appendix}

\section{Note on the influence of the Riemann solver and of the numerical resolution}
\begin{figure}[htb]
  \centering
  \includegraphics[width=9.cm,height=9.cm]{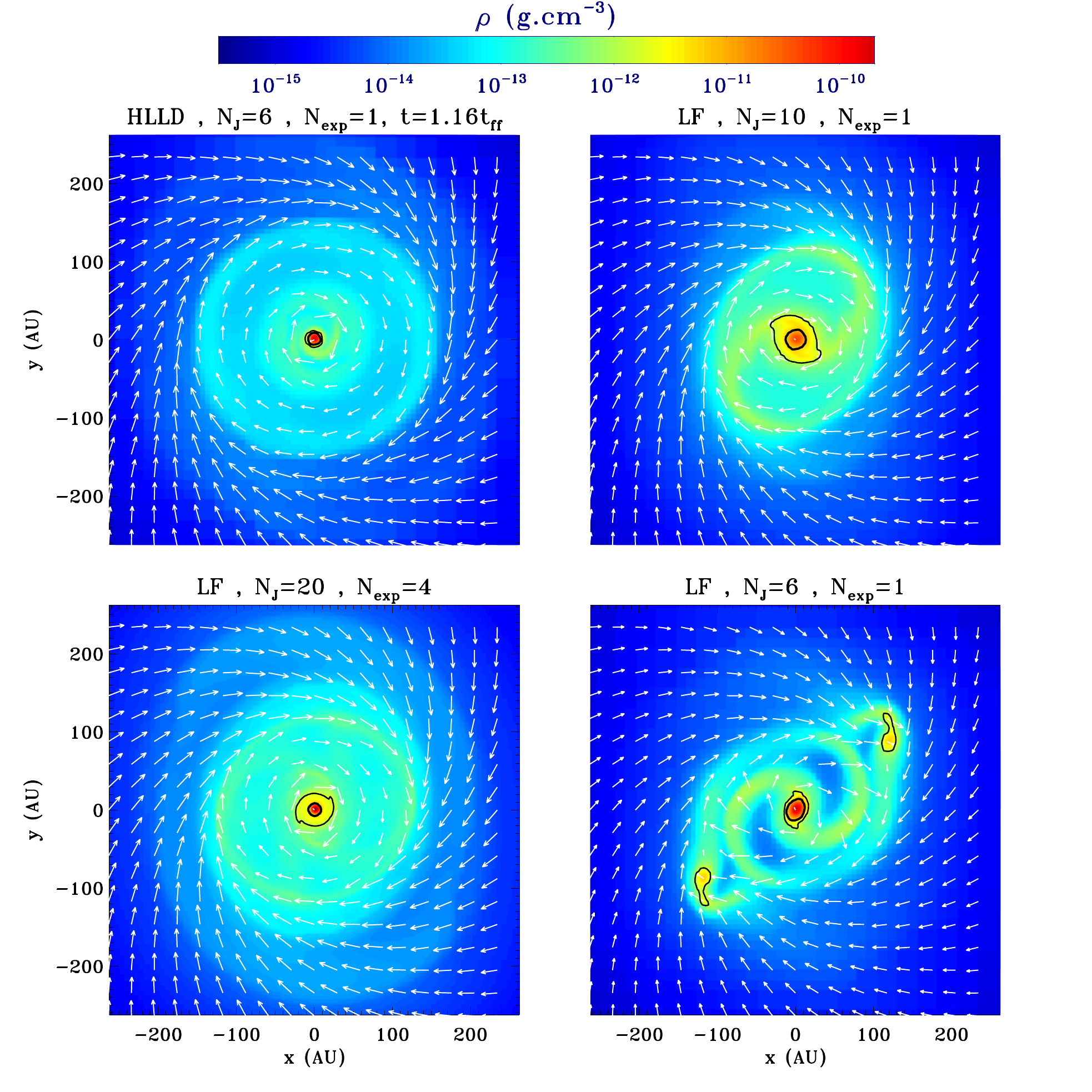}
  \caption{Case $\mu=20$: density maps in the $xy$-plane at time $t\sim1.16\,t_\mathrm{ff}$ for 4 barotropic EOS calculations with HLLD, $N_\mathrm{J}=6$ and $N_\mathrm{exp}=1$; LF, $N_\mathrm{J}=20$ and $N_\mathrm{exp}=4$; LF, $N_\mathrm{J}=10$ and $N_\mathrm{exp}=1$;  LF, $N_\mathrm{J}=6$ and $N_\mathrm{exp}=1$. }
\label{mu20_res}
\end{figure}

In this appendix, we present a short convergence study to investigate the effect of both, the Riemann solver and the numerical resolution, 
on the fragmentation process. We show in section \ref{mu20} that the use of the LF Riemann solver leads to inaccurate results. 
We perform calculations using the barotropic EOS, and the HLLD or the LF solvers, at various numerical resolution. The latter is detemined by 2 parameters: $N_\mathrm{J}$, the number of points per Jeans length, and $N_\mathrm{exp}$, that gives the extent of the mesh around a refined cell (if a cell is flagged for refinement, then the $N_\mathrm{exp}$ cells around the flagged cell, in each direction, will also be refined). A large $N_\mathrm{exp}$ gives a smoother transition between the levels of the AMR grid. All calculations presented in section \ref{mu20} have been run with $N_\mathrm{exp}=4$.
We report results of 4 calculations, run with: HLLD, $N_\mathrm{J}=6$ and $N_\mathrm{exp}=1$; LF, $N_\mathrm{J}=20$ and $N_\mathrm{exp}=4$; LF, $N_\mathrm{J}=10$ and $N_\mathrm{exp}=1$;  LF, $N_\mathrm{J}=6$ and $N_\mathrm{exp}=1$.  

Figure \ref{mu20_res} portrays the density maps in the equatorial plane for the 4 calculations, at time $t\sim1.16\,t_\mathrm{ff}$. 
The calculations performed with the LF solver, $N_\mathrm{J}=6$ and $N_\mathrm{exp}=1$ clearly diverge from the other ones, since it fragments. The lack of resolution clearly induces inaccurate fragmentation. The three other  calculations are qualitatively similar (no fragmentation), and in accordance with  \cite{Hennebelle_Teyssier_2008} results (obtained using a Roe type Riemann solver) . This indicates that the numerical resolution has to be enhanced with the LF solver in order to avoid spurious effects due to the diffusivity of the solver. 

Eventually, note that these differences are more important in the case of FLD calculations, since radiative transfer can have two opposite  effects, depending on the strength of the magnetic braking. If the magnetic braking is not significant (with a bad resolution and a diffuse Riemann solver), the material at the centrifugal barrier cools down and then fragments. On the other hand, when the magnetic breaking is efficient (with HLLD or a high resolution), the infall velocity and consequently, the accretion luminosity, become larger, which prevents fragmentation to occur (material looses angular momentum and heats up). 

\begin{figure*}[hbt]
  \centering
  \includegraphics[width=15cm,height=9.3cm]{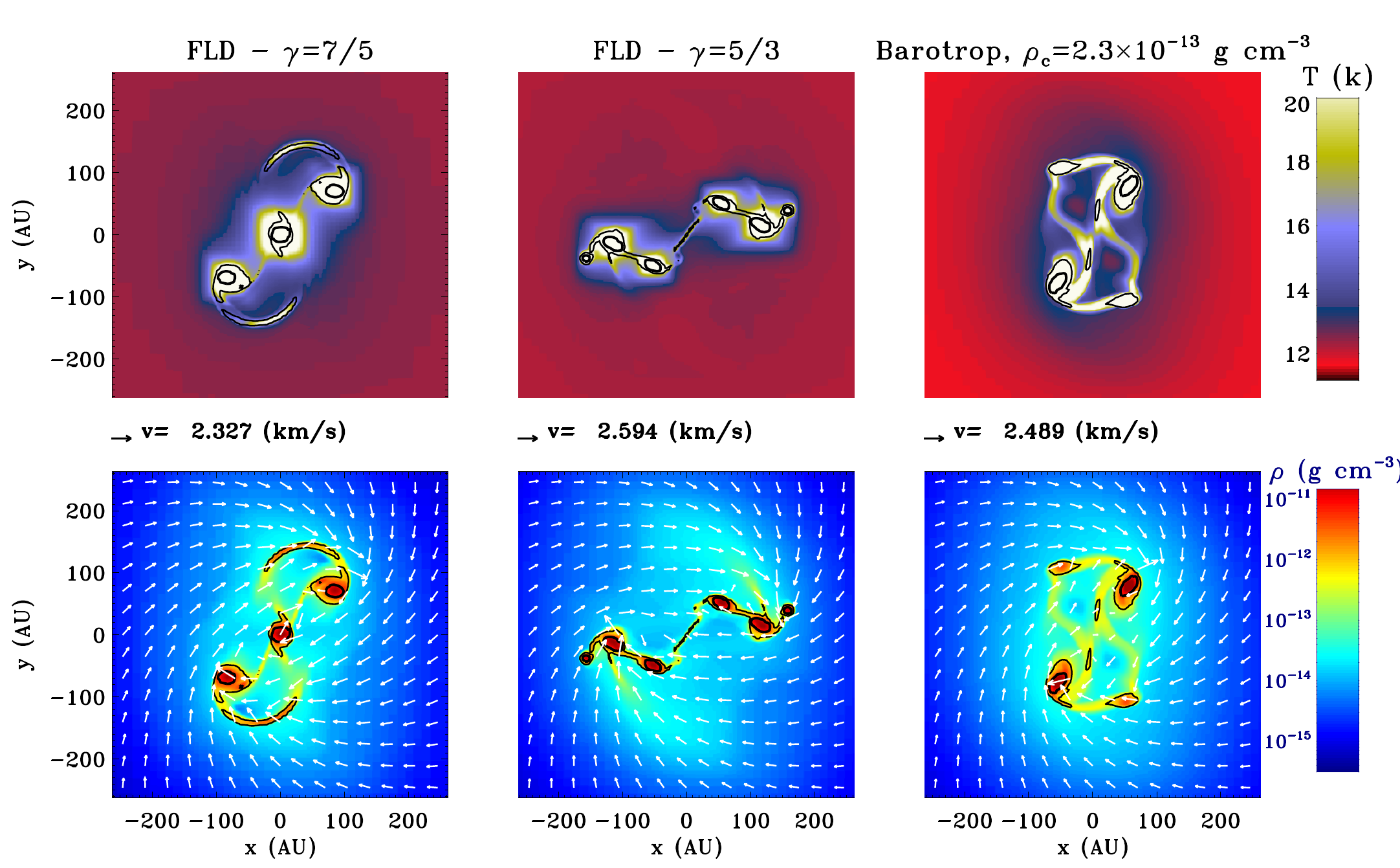}
\caption{Density and temperature maps in the equatorial plane at time t$=38.3$ kyr ($1.15  t_\mathrm{ff}$), for the 3 unmagnetized cases: FLD approximation and $\gamma=7/5$ ({\it left}), the FLD approximation and $\gamma=5/3$ ({\it middle}) and with a barotropic EOS using $\rho_\mathrm{ad}=2.3\times10^{-13}$ g cm$^{-3}$ and $\gamma=5/3$ ({\it right}). Velocity vectors are over-plotted on the density maps}
\label{a037_fld_baro}
\end{figure*}

\section{Case $\mu=1000$}

For this quasi-hydro case, the initial dense core is highly gravitationally unstable. We have performed three types of calculations with the LF Riemann solver: one with the barotropic EOS and $\gamma=5/3$, and two with the FLD and different  adiabatic exponents ($\gamma=5/3$ and $\gamma=7/5$). The value $\gamma=7/5$ is more appropriate when $T> 100$ K, when the rotational degrees of freedom of H$_2$ are excited
\citep[e.g.][]{Machida_et_al_2008}. Using $\gamma=7/5$ yields cooler adiabatic cores than with $\gamma=5/3$ at the same density. Higher densities and temperatures in the core are thus reached more rapidly. 

Figure \ref{a037_fld_baro} portrays density and temperature maps in the equatorial plane for the three aforementioned calculations at a time $t=1.15$ $t_\mathrm{ff}$. Temperature maps range from 11.4 K to 20 K. As pointed out in \cite{Commercon_2008}, the horizon of predictability is very short when the initial thermal support is low. Calculations do not show a convergence to the same fragmentation pattern, since we do not integrate the same system equations. The FLD calculations yield different fragmentation modes for the two different $\gamma$'s: with $\gamma=7/5$, we get a central object and two satellites (4 at later times), whereas with $\gamma=5/3$, we get a ring fragmentation pattern, with orbiting fragments linked by a bar. The barotropic EOS calculation does not produce a central object.
From the temperature map of the FLD case with $\gamma=7/5$, we see that the central region is quite hot. Fragments are more compressed and they radiate more energy. Fragments are also formed faster than the time required for the radiative feedback to become significant (beginning of the second collapse).

\end{appendix}
\end{document}